# DECODING MORAL JUDGEMENT FROM TEXT: A PILOT STUDY

Diana E. Gherman[1], Thorsten O. Zander[1]

[1] Brandenburg University of Technology Cottbus–Senftenberg, Germany

E-mail: diana.gherman@b-tu.de

## ABSTRACT

Moral judgement is a complex human reaction that engages cognitive and emotional dimensions. While some of the morality neural correlates are known, it is currently unclear if we can detect moral violation at a single-trial level. In a pilot study, here we explore the feasibility of moral judgement decoding from text stimuli with passive brain-computer interfaces. For effective moral judgement elicitation, we use video-audio affective priming prior to text stimuli presentation and attribute the text to moral agents. Our results show that further efforts are necessary to achieve reliable classification between moral congruency vs. incongruency states. We obtain good accuracy results for neutral vs. morally-charged trials. With this research, we try to pave the way towards neuroadaptive human-computer interaction and more human-compatible large language models (LLMs).

## INTRODUCTION

*Passive BCIs.* Passive brain-computer interfaces (pBCIs) can seamlessly decode mental states from a user's brain activity [1]. Active BCIs require the conscious and intentional modulation of one's brain activity, while reactive BCIs make use of external stimuli such as flickering lights to evoke a desired reaction [2]. Meanwhile, pBCIs operate in the background, capturing the spontaneous reactions to specific stimuli in the environment. Most commonly, electroencephalography (EEG) signals are collected and used for mental state classification. Once decoded, pBCIs can provide this real-time information to a computer that can then adapt its outputs to cater to individual needs and preferences. This new form of interaction has previously been described as neuroadaptive [3]. Thus, pBCIs could upgrade human-computer interaction (HCI) to a more natural, fluid type of communication that can be employed in various fields. The potential for safer and more efficient occupational environments through neuroadaptivity has been shown for driving [4], aviation [5] and medicine [6], but also for leisure activities such as gaming [7]. Among others, cognitive states like workload [8], error-perception [9] and surprise [10] have been successfully decoded with pBCI. While extensive research has been done to explore average EEG correlates of emotions, there are relatively few studies that demonstrate robust capabilities for emotional state detection at a single trial level [11,12]. The most common types of features used for emotion classification are event-related potentials (ERPs), frontal EEG asymmetry and event-related desynchronization / synchronization [13]. To investigate single-trial emotion detection from ERPs, a recent study combined workload and stress detection in a social evaluation context [14]. Using a cross-subject classification technique with transfer learning, stress vs. relaxation levels were detected with an average accuracy of over 80%. Single-trial classification of emotion based on ERPs was also achieved for different levels of valence and arousal with a definite advantage for arousal discrimination in [15] and [16]. Another study using EEG recorded while participants were watching music videos managed high classification accuracies for stress levels by using entropy-based features [17]. Our study proposes exploring how well pBCI systems can perform in classifying a specific type of emotion, moral emotion [18]. According to the well-known arousal-valence dimension model of emotions [19], moral violations could evoke high arousal and negative valence emotions such as anger or disgust [20,21]. In contrast, congruent moral stimuli could be associated with low arousal and positive valence. In this investigation, we try to decode moral emotions with pBCI through moral judgements.

*Moral judgement.* We operationalize here moral judgement as the degree of agreement or disagreement to morally-charged contexts. Moral judgement is a complex human reaction that can include both a cognitive and emotional dimension [22,23]. As an automatic and emotional response, moral judgement can be triggered at an unconscious, intuition-based level, determined by a combination of factors such as personality, culture or motivation [24,25] and is associated with deeper structures of the brain [26]. On the other hand, especially when explicit moral reasoning is required, cognitive functions such as inhibition, cognitive conflict, memory and theory of mind processes are engaged and different prefrontal cortical areas become more active [27,28]. A morally-charged stimulus can either resonate with or challenge an individual's moral perspective, thereby evoking a meaningful moral reaction. This depends on the congruency moral stance with one's personal values and experience with a particular topic. This reaction can be recorded with brain imaging methods such as EEG and potentially decoded with pBCI. While some EEG studies looked at the signal patterns associated with neutral, positive, and negative moral judgements, there has not been much work investigating the feasibility of single-trial moral judgement detection for text stimuli [29]. In [30], 90 morally consistent and inconsistent statements were presented to pre-selected groups consisting of Christian and non-Christian male participants while recording their electroencephalography (EEG) data. The statements were displayed one word at a time, with the final word of each determining the overall moral meaning. In reaction to these key words, a small N400 event related-potential (ERP) was found for morally-incongruent words. Also, a late positive potential (LPP) was found around 500-600 ms. The congruency of the moral words was determined based on participants' religiosity for relevant topics (e.g. "I think euthanasia is acceptable/unacceptable"). Another similar study [31] used morally acceptable or unacceptable statements (aligned or misaligned with social norms) presented word by word to elicit moral agreement or disagreement. They also found an LPP around the fronto-parietal region in the case of unacceptable statements. A more recent study that used a multivariate pattern classification (MVPA) showed that agreement or disagreement to morally-charged statements (e.g. "Wars are acceptable / unacceptable") could be predicted from 180ms following the critical ending words, based on the approval or disapproval with these statements indicated via button presses ("yes" and "no") [32]. Moral attitudes regarding particular topics are acquired throughout one's life and are strongly correlated with views and values assimilated within family, society, and personal experiences. The context in which statements appear is also important in eliciting corresponding moral reactions. Previous studies have shown that negative emotion can that trigger a signaling mechanism, making moral situations more salient [22]. Thus, a realistic emotional context used as an affective priming for the textual stimulus could significantly help in this elicitation, as compared to passive statements devoid of context [33,34]. This might be especially relevant for single trial detection. Also, existing theories on effective emotion elicitation attest to the importance of constructing agents for moral assessments to be attributed to, which also improve the elicitation of moral reactions, making the experience more relatable and impactful [35,36]. In this paper, we investigate the feasibility of moral judgement decoding with pBCI for morally-charged statements presented following affective priming represented by emotional videos on specific topics. Previous work has identified video-based stimuli with audios to be considerably more efficient in emotion elicitation, as they are more realistic [37] and produce the highest number of statistically significant features [38]. While most studies that used affective priming in the context of moral judgement assessment so far have used text-based priming, we explore the use of videos with audio here. In light of an increasingly digitized world and advanced artificial intelligence systems (AI) such as large language models (LLMs) [39], successful real-time decoding of moral judgement could open a new realm of possibilities for better and more human-compatible HCI through neuroadaptivity.

MATERIALS AND METHODS

*Participants* This pilot study included 3 participants (2 males, and 1 female) with a mean age of 31 years. The experimental procedure was approved by the Research Ethics Committee of the Brandenburg University of Technology Cottbus-Senftenberg (ID: EK2024-03).

*EEG recording.* Their EEG data was recorded using 64 active actiCAP slim gel electrodes (Brain Products GmbH, Gilching, Germany), according to the 10-20 international system. The signal was sampled at 500Hz. The Lab Streaming Layer (LSL) [40] was used to synchronize the channel streams.

*Experiment overview.* The task involved watching videos and reading statements related to 4 social justice issues: immigration, racial discrimination, sexism, and homosexuality. Sixteen videos were presented in a random order, followed by 10 randomized statements (5 morally agreeable/congruent and 5 morally disagreeable/incongruent). The utilized videos were collected directly from YouTube or complied together using sequences from a longer Youtube video, such that each video lasted approximately 1 minute. They represented a segment from real TV or media news and they were generally found on channels of multimedia news organizations. Each video included audio as well.

After each visualisation, the participants would read an instruction informing them the upcoming statements would be comments left under the respective video by people on the internet. Thus we are framing here strangers on the internet as moral agents responsible for their actions, here agreeable or disagreeable statements. In reality, statements were created by experimenters with the help of the publicly available LLM, ChatGPT. The written statements presented word by word, in a Rapid Serial Visualization Presentation (RSVP) manner [41] with an Optimal Recognition Point (ORP) alignment of the words [42]. Corresponding with the ORP position, the aligned letters in each word were presented in red, while the rest of the letters remained black. While pre-defined labels were set for these key words, the participants also indicated their agreement, disagreement, or uncertainty regarding the read statement by using keyboard buttons (left arrow for *agree*, downward arrow for *uncertain* and right arrow for *disagree*). The position of the agreement level buttons on the screen corresponded with the position of the response keyboard buttons. The ending, key word of each sentence determined the entire moral stance of the sentence and represented the events used for classification. In total, there were 160 statements and therefore, 160 key words in the task: 80 morally-congruent and 80 morally-incongruent. Following, we will illustrate one video-statements example. One of the videos included in the study was a short news piece on the persecution and abuse gay people experience in Uganda due to strict anti-homosexuality laws. In this, there are depictions of people expressing their disappointment and fear regarding these laws and sequences of politicians communicating morally controversial statements such as: "We are going to reinforce the law enforcement officers to make sure that homosexuals have to space in Uganda." After this video the sentences in Tab. 1 were presented in a randomized, word by word manner. The speed of the word presentation differed based on the character length of each non-key word, with 700ms base time and 20ms added for each character besides the first one. For instance, the word *must* was presented for 760ms. The ending, morally-charged words were all presented for 1500ms. The task lasted one hour.

Table 1. Example of morally congruent and incongruent statements

| Congruent |
|---|
| 1. Uganda's laws for gays should be **humane**. |
| 2. Gay people in Uganda deserve **freedom**. |
| 3. Gay people in Uganda should be treated with **dignity**. |
| 4. Gay Ugandan citizens merit more **respect**. |
| 5. Equality for gay individuals in Uganda is **essential**. |
| Incongruent |
| 6. In Uganda, laws regarding gays must be **harsh**. |
| 7. Uganda's gay people deserve **prison**. |
| 8. Gay Ugandans should be shown **disrespect**. |
| 9. Ugandan gays merit more **punishment**. |
| 10. For Uganda's gays, equality is **unacceptable**. |

*Classification method.* The classification was performed offline, using MATLAB R2022a (The Mathworks, Inc., Natick, MA, USA) and BCILAB 1.4-devel [43]. Responses that did not align with the predefined classes (*congruent* vs. *incongruent*) were excluded from the classification. Thus, in the sentence "Gay people in Uganda deserve freedom." the pre-defined label for the word *freedom* was *congruent*. If the participants pressed on the "disagree" or *uncertain* buttons instead, this trial was excluded from the classification. We also explored the classification of *moral* (congruent and incongruent moral combined trials) vs. *neutral* trials. The neutral trials were categorized based on list of 86 words that appeared within sentences. Examples of neutral words include: "eventually, ultimately, casual, concept, idea, fact". A windowed means approach [44] was used for the feature extraction. The data was bandpass-filtered between 0.1 and 15 Hz. Regularized linear discriminant analysis (LDA) with a (5x5)-fold cross-validation was used for the classification of congruent vs. incongruent trials and moral vs. neutral trials. Epochs of 1 second were extracted with a start time at stimulus onset (key word presentation). To capture potential N400 and LPP effects, we explored two sets of 50 ms time windows in which amplitude is averaged. One set of time windows we used were between 300 and 600 ms after the stimulus, with 6 consecutive time windows. The second set of time windows were set between 400ms and 1000 ms, with 12 consecutive time windows.

RESULTS

The average classification results on congruent vs. incongruent classes (*CvsI*) and neutral vs. moral (*NvsM*) for both sets of time can be seen in Tab. 2. Only one participant reached classifier significance for the 400-1000 set, with an accuracy of 65% and chance level at 57% (not shown in Tab. 2). In contrast, all classifiers for both time window sets reached significance for the neutral vs. moral trials. Averaged ERP potentials for channels Fz and Cz were obtained for both types of classes after independent component analysis (ICA) and non-brain component removal. ERPs for morally congruent vs. incongruent trials are illustrated in Fig. 1 and ERPs for morally-charged vs. neutral trials are illustrated in Fig. 2.

Table 2. Classification results for congruent vs. incongruent (*CvsI*) and neutral vs. moral (*NvsM*) trials

| Time windows | TP (%) | TN (%) | Accuracy (%) |
|---|---|---|---|
| 300 - 600 CvsI / NvsM | 49 / 83 | 52 / 69 | 50 / 78 |
| 400 - 1000 CvsI / NvsM | 59 / 80 | 54 / 72 | 57 / 77 |

*TP = True positives (incongruent); TN = True negatives (incongruent)*

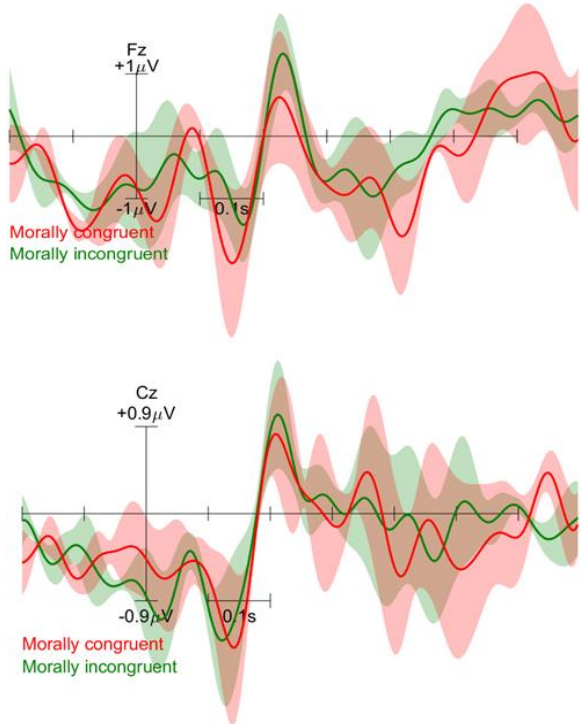
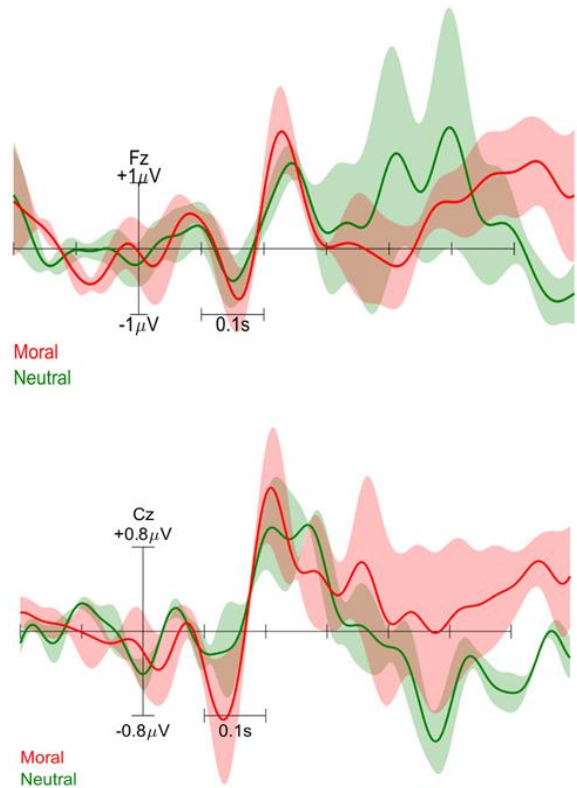

Figure 1. Average ERP potentials for morally congruent and incongruent words.

Figure 2 Average ERP potentials for moral and neutral words.

DISCUSSION

While decoding accuracy for morally congruent and incongruent trials was not successful with this simple approach, we could observe good decoding accuracies for neutral vs. morally-charged words. This was also reflected in the grand-average ERP. Our results are not entirely surprising, given the difficulty of emotion detection from EEG at a single-trial level [12] and the complexity of moral emotions. A recent pBCI investigation [29] also found chance-level results when looking at the potential of single-trial detection for morally acceptable and objectionable trials on data collected in [30] and [31]. However, we found good performance classification for neutral vs. morally-charged trials. We postulate that while the chosen moral words are relevant enough to produce genuine reactions in comparison to neutral stimuli, the current feature extraction and classification approach might need improvements to better capture potential signal differences between morally congruent and incongruent trials. Encouraging results come from recent studies that explored more sophisticated algorithms and feature extraction methods for emotion detection [17,45]. Another way we plan to improve our results in a larger study is to only include participants that align with a certain profile, such that we can ensure they hold clear moral stances towards the topics. Previous studies have identified the importance of moral attitude strength for effective moral emotion elicitation [46] and the corresponding impact on neural signals [23]. More specifically, we will include questionnaires meant to assess the participants' attitudes towards sexism [47], immigration [48], racism [49] and homosexuality [50]. Hence, only participants who highly agree with immigration and homosexuality and highly disagree with sexism and racism will be invited to the study. Successful real-time decoding of mental states in reaction to written stimuli could transform human-computer communication in the context of LLMs. For instance, training of LLM could benefit from replacing or augmenting human explicit feedback in Reinforcement Learning with Human Feedback (RLHF) [51,52] with neural-based implicit feedback [53], potentially offering new solutions for a better synergy between humans and machines.

CONCLUSION

In this pilot investigation, we looked at the feasibility of single-trial detection of moral judgement from text after video-based affective priming. Our work offers insights into the neural correlates of moral judgement, as well as ideas for classification improvement for a study that includes more participants and better-suited participant profiles.